\documentstyle[12pt,twoside,fleqn,espcrc1]{article}

\begin{document}

\newcommand{\hf}{h(\rho)}
\newcommand{\ho}{{\hbar\omega}}
\newcommand{\jz}{{J_z}}
\newcommand{\oj}{{\omega\jz}}
\newcommand{\id}{{\cal J}_2}

\vglue 10mm
\noindent {\Large Microscopic approach to collective motion}

\vskip 1em
\noindent P. Bonche$^{\rm a}$, E.~Chabanat$^{\rm b}$, B.Q. Chen$^{\rm c}$,
          J. Dobaczewski$^{\rm d}$, H. Flocard$^{\rm e}$, B. Gall$^{\rm e}$,
          P.H. Heenen\footnote[1]{Directeur de Recherches FNRS}$^{\rm ,\; f}$,
          J. Meyer$^{\rm b}$, N. Tajima$^{\rm g}$ and M.S. Weiss$^{\rm h}$
\vskip 1em
\noindent $^{\rm a}${SPhT\footnote[2]{Laboratoire de la Direction des
          Sciences de la Mati\`ere du CEA}, C.E. Saclay, 91191 Gif sur Yvette
          Cedex, FRANCE}

\vskip 1em
\noindent $^{\rm b}$IPN Lyon, CNRS-IN2P3 / Universit\'e Lyon 1,
  69622 Villeurbanne Cedex, France

\vskip 1em
\noindent $^{\rm c}$W.K. Kellog Radiation Laboratory,
 Caltech, Pa\-sa\-dena, California 91125, USA

\vskip 1em
\noindent $^{\rm d}$Institute of Theoretical Physics,
  Warsaw University, Ho\.za 69, PL-00681 Warsaw, Poland

\vskip 1em
\noindent $^{\rm e}$Division de Physique Th\'eorique\footnote[3]{Unit\'e de
  recherche   des Universit\'es Paris XI et Paris VI, associ\'ee au CNRS},
  I.P.N., 91406 Orsay Cedex, France

\vskip 1em
\noindent $^{\rm f}$Service de Physique Nucl\'eaire Th\'eorique,
  U.L.B., CP229, 1050 Brussels, Belgium

\vskip 1em
\noindent $^{\rm g}$Institute of Physics, University of Tokyo, Komaba,
  Meguroku, Tokyo 153, Japan

\vskip 1em\noindent$^{\rm h}$Physics Department, LLNL, Livermore,
   California 94550, USA

\vspace {2em}

An overview of a microscopic framework based on the Hartree-Fock description
of the mean field is presented which, starting from an effective interaction
allows a description of collective motions.
A study of the isotope shifts in the Pb region illustrates the importance of
the effective interactions and points to their limitations.
Such forces should be improved in order to achieve a better description of
nuclei properties especially with the coming availability of exotic beam
facilities.
The coupling of collective quadrupole and octupole degrees of freedom in
$^{194}$Pb is analyzed within the Generator Coordinate Method, which
represents a step beyond the simple mean-field approximation.
As a last example, we present a study of nuclear rotations.
First we discuss results on superdeformed rotating bands in $^{192}$Hg,
$^{194}$Hg and $^{194}$Pb obtained without including a treatment of pairing
correlations.
Preliminary calculations are also presented with these correlations included
as well as an approximate projection on nucleon number.

\section{THEORETICAL OUTLINE}

Two types of basic ingredients can be found in most mean field calculations;
first a two-body nucleon-nucleon effective interaction, such as a Skyrme-type
one, complemented by a pairing interaction.
The many body wave function representing the nucleus is approximated by a
Slater determinant or a BCS state.
Then a minimization of its energy yields the Hartree-Fock (HF) or mean field
equations.
These equations are non linear as the mean potential depends upon the solution
through different densities (mass, kinetic energy...) in addition to the
explicit density dependent terms which appear due to the density dependence
of the effective interaction.
The numerical solution is performed via an iterative procedure from an
initial guess of the solution.
In practical calculations, the BCS equations for the pairing interaction are
solved at each HF iteration so that pairing correlations are included
self-consistently into the mean field and not only in first order
perturbation theory as when pairing is treated as a residual interaction.
Since the HF equations result from a variational principle, the solution is
an energy extremum, the absolute minimum being obtained when the initial
conditions are properly chosen.

After the choice of the interaction comes the second key ingredient
to describe collective motions in mean field calculations, namely
a set of constraining operators which are included to explore these
collective degrees of freedom.
The minimization of this modified energy functional yields the constrained
Hartree-Fock (CHF) equations~\cite{Zrpaper,PB89,Isthmus}.
As before, their solution gives a Slater determinant representing the many
body wave function of the nucleus, subject to the chosen constraints.
The choice of the constraining operator (or operators) depends upon
the physics.
Degrees of freedom associated to the shape of the nucleus are
studied mostly by means of quadrupole constraints~\cite{Zrpaper}.
As an example, we will present below an analysis~\cite{Mey93} of the
quadrupole-octupole coupling in $^{194}$Pb in which both octupole and
quadrupole operators have been added to the HF mean field.
Similarly, the fission path of heavier nuclei can be studied with a quadrupole
constraint together with a hexadecapole one.
In every cases, the resulting shape of the nucleus is {\it optimized} with
respect to all other unconstrained operators, e.g. of other multipolarities,
due to the variational nature of the CHF equations.

To study rotating nuclei, one generally adds a constraint on the projection
of the angular momentum onto the appropriate axis~\cite{Mg24}.
To generate rotational bands in superdeformed (SD) nuclei, a constraint on
the quadrupole deformation is also introduced to construct the SD secondary
minima on which one wants to build a rotational band.
In the Hg region, where such secondary minima do exist at zero angular
momentum, the quadrupole constraint can be released when the nucleus is
cranked up.
Whereas in other regions of the mass table, no such minimum exists and one must
keep the quadrupole constraint until the angular momentum is high enough
to stabilize the deformation.

In all these cases, nuclei are described by intrinsic states
which usually break several symmetries.
In particular they are no longer eigenstates of the angular momentum operators,
nor of the particle number operators when pairing correlations are included.
Only the mean value of these operators is accounted for through the
constraining fields.

We have considered two methods which allow to get a better description of the
nuclear properties beyond the mean field approximation.
The first one is the Generator Coordinate Method (GCM) which provides
collective wave functions associated to a given mode~\cite{Hg92}.
Let us make three comments,
i) in all the cases we have studied until now, this method reduces to a
diagonalization of the residual interaction in the configuration space built
on a finite set of constrained intrinsic HF states, with the additional
technical difficulty that these states are not orthogonal.
ii) these HF states which generate the GCM basis are bound by the collective
potential: we never considered situations involving scattering states.
The collective wave functions which we calculate represent bound states
within the collective well.
iii) finally, as the generating states are built from the same (Skyrme)
interaction which is used in the GCM calculation, the dimension of the
GCM basis remains rather small.
Typically 30 to 40 CHF states at most are usually required to describe
quadrupole deformations from oblate to prolate SD shapes and even
hyperdeformed (HD) ones for axially symmetric configurations.
If one were to use another set of generating states, this number
would be much larger.
Somehow this number is minimal when the generating states incorporate
most of the mean field effects of the two body interaction used in the GCM.

The second method realizes the restoration of broken symmetries through
projection.
This can be done in two ways, the simplest method consists in projecting
the HF solution onto good quantum numbers.
Along this line, particle number projection has been done in some
cases~\cite{Proj}.
Angular momentum projection has also been performed in an approximate manner
by an appropriate use of the specific symmetries of the triaxial GCM
in the Sr region~\cite{Srpaper}.
A second and more accurate way involves a projection before variation.
Since it is rather difficult to implement, we have resorted the Lipkin-Nogami
(LN) prescription~\cite{Lipnog,Pb1s} which emulates an approximate projection
up to the second order in $\hat{N}-<\hat{N}>$ onto the particle number
operator $\hat{N}$ before variation.

In what follows, we illustrate the main points which have been listed above.
First we present a study~\cite{Taj93} of the isotope shifts of Pb nuclei to
emphasize the importance of the choice of the effective interaction and
discuss their range of validity.
Then we present an analysis of the quadrupole-octupole modes in $^{194}$Pb
together with a GCM calculation of the corresponding collective wave
functions~\cite{Mey93}.
Finally a study of identical bands in the Hg-Pb region is given
without pairing correlations~\cite{Bing}.
Preliminary calculations with these correlations included are also presented
and the effect of restoring symmetry breaking is discussed in the case of an
approximate projection onto good particle number via the LN prescription.
The choice of these topic aims at covering the different aspects of mean
field calculations, namely:
definition of an effective force, choice of the collective variables (or
equivalently constraining operators), exploration of collective correlations
beyond the mean-field, and finally partial restoration of symmetry breaking.

\section{ISOTOPE SHIFTS OF THE PB NUCLEI}

The isotope shifts of atomic levels provide a very good testing ground
for nuclear forces.
They are experimentally measured with a high accuracy.
 From their values, the variation of root-mean-square charge radii
can be obtained with little ambiguity.
The Pb isotope sequence is particularly interesting as the $^{208}$Pb
doubly-magic nucleus is one of the anchor points in the parametrization
of effective interactions for mean field calculations.

The behavior of the r.m.s. radii presents too features.
One is a kink at the $^{208}$Pb shell closure when they are drawn
as a function of the mass number.
The other one is an odd-even staggering.
These features are not specific to the Pb isotopic chain and are observed
in other region of the mass table~\cite{Isoexp}.

\subsection{Mean field calculations}

We have calculated the proton radii with three different Skyrme forces.
First we used SIII~\cite{SIII} which is one of the most successful Skyrme
force for spectroscopic properties.
However it has a rather large incompressibility modulus ($K_{\infty}=355$ MeV)
and predicts too large fission barriers.
Then we used  SkM$^{\ast}$~\cite{Skmast} and SGII~\cite{Sg2} which are derived
from the SkM force~\cite{SkM}.
The latter is intended to give correct energies for isoscalar $E0$ and
isovector $E1$ giant resonances whereas SkM$^{\ast}$ improves upon surfaces
tension to allow studies of fissioning nuclei.
Both SkM$^{\ast}$ and SGII have a smaller value of $K_{\infty}$ than SIII.

\begin{figure}[htb]
\begin{minipage}[t]{160mm}
\vskip 105mm
\caption{Proton (circle) and neutron (squares) mean-square radii of
Pb isotopes for different Skyrme forces.
The experimental points are connected by a dotted line.}
\label{F:pbr}
\end{minipage}
\end{figure}

We define changes in mean square radii with respect to $^{208}$Pb as
    $\Delta r^2 (A)$
    $\equiv$
    $r_2 (A)$ $-$
    $r_2 (208)$,
where $r_2 (A)$ is the mean square radius of the isotope of mass $A$
whose wave function is calculated within HFBCS.
Fig.~\ref{F:pbr} shows calculated and experimental proton $\Delta r^2 (A)$
as a function of the mass number.
To all these quantities we have subtracted the change in mean square radii
of a liquid drop,
    $\Delta r_{\rm LD}^2 (A)$
    $\equiv$
    $r_{\rm LD}^2 (A)$ $-$
    $r_{\rm LD}^2 (208)$,
with
    $r_{\rm LD}^2$ =
    $\frac{3}{5} r_0^2 A^{2/3}$
and
    $r_0$ = 1.2 fm.
This subtraction emphasizes deviations from any smooth trend.

If the agreement with experiment is excellent for SkM$^{\ast}$ and SGII
for neutron deficient isotopes, all calculations fail to reproduce
the abrupt change in slope at $A=208$.
As the slope is almost linear both sides of $^{208}$Pb, we characterize this
change in slope by its {\sl kink} defined from $^{194}$Pb, $^{208}$Pb,
and $^{214}$Pb as
$k$ $\equiv$
$\frac{1}{6} \Delta r^2 (214)$ $-$
$\frac{1}{14} \Delta r^2 (194)$.
The values of $k$ are given in table~\ref{T:kink} with those of $\Delta r^2$,
both for protons and neutrons.
The size of the kink is larger for neutrons, which is natural as the proton
radius can only change through the p-n attractive force when one varies the
neutron number for a given charge.
A stronger p-n force would result in a larger kink,
however, as the neutron kink is smaller than the proton experimental one,
it is unlikely that the sole modification of the symmetry properties of
the force will resolve the discrepancy.
Let us also note that the magnitude of $k$ is not directly related
to the incompressibility modulus.
The SkM$^{\ast}$ and SGII forces give very different values of $k$,
although they have the same value of $K_{\infty}$.
The kink obtained for SIII lies in between whereas its $K_{\infty}$
is much larger.
Finally, Hartree Fock Bogoboliubov (HFB) calculations have also been
done~\cite{Jfb} with the D1S interaction~\cite{D1,D1S}.
They produce a kink of 0.12 fm$^2$ which is as small as that of SIII.

\begin{table}
\begin{minipage}{118mm}
\caption{Differences of proton and neutron radii (fm$^2$) from those of
$^{208}$Pb calculated with three types of Skyrme forces.}
\label{T:kink}
\begin{tabular}{ccccccc}
\hline
force &
\multicolumn{2}{c}{$\Delta r_{\rm p}^2$} &
$k_{\rm p}$ &
\multicolumn{2}{c}{$\Delta r_{\rm n}^2$} &
$k_{\rm n}$ \\
 & ($^{194}$Pb) & ($^{214}$Pb) &  & ($^{194}$Pb) & ($^{214}$Pb) &  \\
\hline
SkM$^{\ast}$ & $-0.7573$ &  0.3545 &  0.0050 & $-1.5931$ &  0.8530 &  0.0284 \\
SIII         & $-0.8680$ &  0.4481 &  0.0127 & $-1.4939$ &  0.8335 &  0.0322 \\
SGII         & $-0.7448$ &  0.4576 &  0.0231 & $-1.4589$ &  0.8843 &  0.0432 \\
exp          & $-0.6830$ & 0.6099  &  0.0529 &           &         &         \\
\hline
\end{tabular}
\end{minipage}
\end{table}

It is important in a study of the isotope shifts based on an effective force
to verify the quality of that force for the calculated binding energies.
With $^{208}$Pb as a reference point, fig.~\ref{F:pbe} shows the
predicted binding energy difference,
    $\Delta E (A)$
    $\equiv$
    $E (A)$ $-$
    $E (208)$,
to which we have subtracted the equivalent experimental quantity.
The analysis of this figure shows that $\Delta E (A)$ strongly deviates
from zero.
This implies that none of these three forces can be considered to have
correct symmetry energy properties.

\begin{figure}[htb]
\begin{minipage}[t]{125mm}
\vskip 80mm
\caption{Error of calculated masses of Pb isotopes with SGII (dashed line),
SkM$^\ast$ (solid line), and SIII (squares).
Differences are plotted with respect to $^{208}$Pb.}
\label{F:pbe}
\end{minipage}
\end{figure}

Nevertheless, from the present calculation, it appears to us that SIII is the
best choice among these three forces.
Indeed it gives results comparable in quality as one moves away either side
of $^{208}$Pb.
Had we chosen  $^{194}$Pb as a reference point to calculate $\Delta r^2$ and
$\Delta E$, none of the three forces would have predicted the $^{208}$Pb
binding energy correctly, however SIII alone would have given a remarkable
agreement for $^{214}$Pb, with $\Delta E<100$ keV and $\Delta r^2<0.023$
fm$^2$.

\subsection{Corrections}

To improve upon these mean field calculations, we have calculated different
corrections.
First of all the experimental isotope shifts measure a difference
in charge radii whereas HF calculations give a proton density.
We have then folded the proton density with the charge distribution
of the proton, taking into account effects coming from the neutron
charge distribution and the spin density~\cite{Ma80}.
These effects are small and roughly proportional to the mass number.
They modify slightly the charge radii of the three nuclei we use to calculate
the kink which does not change in a significant manner, less than 1\%

The experimental isotope shifts depend mostly upon the mean square charge
radii, however higher moments of the charge density also enter
in their determination.
The ratio of the contribution of these higher moments to that
of the mean square radius is estimated assuming a uniform charge
distribution~\cite{Tho83,Din87}.
We extracted from our calculations the expectation values of $r^4$
and found that even though the expectation values of both $r^2$ and $r^4$
are overestimated in the sharp density model, their ratio agrees rather well
with our numbers.
It is thus very unlikely that any refinement of the experimental
analysis will provide an explanation for the kink.

The above corrections remain within the scope of the predictions from
a mean field calculation.
We have also investigated possible corrections coming from the admixture
of collective modes into the HFBCS ground states,
extending our calculations beyond the mean field approach.
This is done by utilizing the GCM where the generating variable is chosen
according to the mode we want to study.
In all these GCM calculations, we have used the SkM$^{\ast}$ interaction.
Quadrupole modes were first studied in this way.
The resulting modification is the isotope shifts is very small,
at most 1\% in $^{194}$Pb, even smaller in $^{214}$Pb.
A GCM analysis of the isoscalar breathing mode lead to similar results.

None of the above contributions presents dramatic
changes across the doubly-magic $^{208}$Pb nucleus.
Experimentally, the level scheme shows evidence for a modification
of the octupole properties when crossing the magic shell closure.
For masses smaller than 208, the lowest $3^{-}$ state is similar
to that of $^{208}$Pb although its collectivity decreases
as one removes pairs of neutrons.
For neutron rich isotopes, additional $3^{-}$ states can be constructed
from the excitation of one neutron
from the now populated g$9/2$ shell to the j${15/2}$.
On this basis, we have studied possible contribution to the kink
from coupling to octupole modes.
However, in that case too, the complete GCM calculation did not
significantly modify our mean field results.

In view of the smallness of these corrections, we did not think
that a more complete two- or even three-dimensional GCM calculation,
mixing the three moments together, would change our conclusion.

\subsection{Density-dependent pairing forces}

Instead of the constant G pairing interaction, we use below a more realistic
zero-range $\delta$-pairing force~\cite{Tondeur,Kri90} for the BCS calculation,
together with the SkM$^{\ast}$ nuclear force.
We choose to quench this pairing interaction inside the nucleus,
as is usually the case of Landau-Migdal particle-hole effective
forces~\cite{Landau}.
With this choice, the nucleus tends to expand when the pairing correlation
is switched on because the interaction is stronger at lower densities,
e.g. in the vicinity of the surface.
To achieve this quenching,
we have introduced density dependence in the pairing interaction as follows:
\begin{equation} \label{DltFrc}
V^{\tau}(\vec{r}_1,\sigma_1;\vec{r}_2,\sigma_2)
=V^{\tau}_0 \frac{1-\vec{\sigma}_1 \cdot \vec{\sigma}_2}{4}
\delta(\vec{r}_1 - \vec{r}_2)
f( \frac{\vec{r}_1+\vec{r}_2}{2} ).
\end{equation}
As the density dependence function
$f(\vec{r})$, we choose, for the sake of simplicity,
linear dependence on the nucleon density parameterized by a critical density
$\rho_{\rm c}$,
\begin{equation} \label{DDFct}
f(\vec{r}) = 1 - \frac{\rho (\vec{r}) }{\rho_{\rm c}}.
\end{equation}
Such a density-dependent parametrization have already been proposed
and used in into different physical situations such as the analysis
of pairing in the actinide region~\cite{Cha76} or a more recent study
of pair correlations near the neutron drip line~\cite{BE91}.
In these studies, the $f$ function depends upon the density through a
fractional power.

One can schematically show how this new pairing force will affect the radius.
Assuming a constant density $\rho_0$ within a sharp surface,
the average size of the two-body pairing matrix element is estimated to be,
$\bar{v} = (V_0 / N )\;  \rho\; (1-\rho_0 / \rho_{\rm c} ) $,
which is maximum at $\rho_0 = \rho_{\rm c}/2$.
In the case of a pure $\delta$-force ($\rho_{\rm c}=\infty$),
pairing compresses the nucleus as compared to the HF solution.
If $\rho_{\rm c} = \rho_0$, pairing is turned off inside the nucleus,
it acts as a surface interaction~\cite{Mosko} and the nucleus expands.

\begin{table}
\caption{Neutron and proton mean square radii (in fm$^2$) with
density dependent pairing.}
\label{T:ddkink}
\begin{tabular}{llrccrccr}
\hline
  &
$\rho_{\rm c}$ &
$V_{0}^{\rm n}$ &
\multicolumn{2}{c}{$\Delta r_{\rm p}^2$} & $k_{\rm p}$ &
\multicolumn{2}{c}{$\Delta r_{\rm n}^2$} & $k_{\rm n}$ \\
 & [fm$^{-3}$] & [MeV fm$^3$] &
($^{194}$Pb) & ($^{214}$Pb) &  & ($^{194}$Pb) & ($^{214}$Pb) & \\
\hline
a) &$\infty$&$-230$ &$-0.8948$&0.3097&$-0.0123$&$-1.7440$&0.7930&  0.0076 \\
b) &\multicolumn{2}{r}{---$\;\;\;\;\;g_{\rm n}$=12.5}
                    &$-0.7573$&0.3545&$ 0.0050$&$-1.5931$&0.8530&  0.0284 \\
c) &0.1603  &$-750$ &$-0.6662$&0.3753&$ 0.0150$&$-1.5137$&0.8668&  0.0364 \\
d) &0.1382  &$-1150$&$-0.5165$&0.4454&$ 0.0373$&$-1.3571$&0.9453&  0.0606 \\
e) &0.13    &$-1450$&$-0.4464$&0.4761&$ 0.0475$&$-1.2825$&0.9817&  0.0720 \\
exp&        &       &$-0.6830$&0.6099&$ 0.0529$&         &      &         \\
\hline
\end{tabular}
\end{table}

Table~\ref{T:ddkink} gives the parameters of the neutron pairing force used
in the calculation, no proton pairing is considered.
The set (a) correspond to a pure $\delta$-force, it tends to compress the
nucleus, (b) is our previous constant $G$ pairing calculation for comparison.
The sets (c), (d) and (e) correspond to different choices of $\rho_{\rm c}$.
In case (c), $\rho_{\rm c}$ is the saturation density of SkM$^\ast$, while in
case (d), it is the density of a liquid drop of radius 1.2$A^{1/3}$ fm.
When calculated with SkM$^\ast$, the average interior density of the Pb
isotopes is about 0.158 fm$^{-3}$, which is between (c) and (d).
An even smaller value of $\rho_{\rm c}$, (e), has been used to show the
sensitivity of the kink to this parameter.
In all cases, the strength $V_0$ was adjusted to reproduce the experimental
pairing gaps.

If the value of the kink can be fitted to experiment, while adjusting
$\rho_{\rm c}$, the overall trend of the isotope shift still does not agree
with the data.
The increase of the kink is obtained by raising both branches of the
theoretical curves on fig.~\ref{F:pbr} until they match to the correct
angle, the overall slope however remains the same and thus incorrect.
Indeed this is related to the symmetry-energy property of SkM$^\ast$
which has remained almost unchanged: the total binding energies
do not vary significantly (less than 0.8 MeV)
for $^{194}$Pb and $^{214}$Pb from case (a) to (e).

\subsection{The effective interaction revisited}

In the above section, we have shown that the symmetry properties of most used
Skyrme forces cannot reproduce the overall behavior of the isotope shifts.
This statement holds for the D1S force as well~\cite{Jfb}.
For SkM$^\ast$, we have also shown that an adjustment of the pairing
interaction can resolve part of the discrepancy.
Therefore a more correct procedure should involve first an improvement of the
symmetry property of the force, then a tuning of the pairing force, if need be.

In the usual fitting procedures~\cite{SIII,Tondeur}, the Skyrme parameters
are fitted to a limited set of quantities.
Among them is the saturation property of infinite nuclear matter and the
incompressibility modulus.
One also includes ground state properties of a small set of closed shell
nuclei together with the surface tension and the symmetry energy.
However, these last quantities are known only in the vicinity of the
$\beta$ stability line.
Calculations are then carried on for other nuclei throughout the mass region
to predict their spectroscopic properties until some experimental evidence
shows that new  features have to be incorporated in the parametrization of
the force.

 From this point of view, the symmetry energy property of the force is a local
property: if a nucleus is correctly calculated, neighboring ones will be
calculated with a similar accuracy.
A study of an isotopic chain is exploring the global aspect of the N-Z degree
of freedom, this feature has not been taken into account in the adjustment
of the force.

To improve on the global symmetry energy, we decided to fit a new set of
Skyrme parameters not only to symmetric nuclear matter, but also to the
extreme case of dense neutron matter.
For that purpose we took the UV14 plus UVII equation of state~\cite{Wiringa}
(eos) and the saturation property of symmetric matter as input to the fit.
We have verified that the resulting Skyrme eos does reproduce the
normal density neutron matter eos~\cite{FP}.
The next step in the adjustment is to reproduce the binding energies
of a limited set of spherical nuclei ranging from $^{16}$O to $^{208}$Pb
within the constraints given by the fit of symmetric and neutron
infinite nuclear matter.

For the resulting set of parameters~\cite{Chabana},
the upper left part of fig.~\ref{F:cha} shows the energy difference between
the UV14 plus UVII and the Skyrme eos for neutron matter.
The other inserts correspond to different Skyrme forces.
As one can see, none of the present parametrization reproduces correctly
the neutron matter eos of ref.~\cite{Wiringa},
only our new set of parameters does it.
As these forces have a similar symmetry energy coefficient, this illustrates
that it is only a local measure of the symmetry properties of the force.
The success of our fit shows that it is possible to improve them without
necessarily modifying the analytical structure of the Skyrme force
or adding new terms into it.

Our fit remains very preliminary, as it should be confronted to more nuclear
spectroscopic properties and eventually improved.
Finally, isotope shifts are to be calculated again to test the need
of a density dependence of the pairing interaction.

\begin{figure}[htb]
\begin{minipage}[t]{133mm}
\vskip 130mm
\caption{Energy differences between various Skyrme and the UV14+UVII equation
of state for neutron matter as a function of the neutron density.}
\label{F:cha}
\end{minipage}
\end{figure}

\section{QUADRUPOLE-OCTUPOLE COUPLING}

In the previous microscopic analysis of quadrupole and octupole modes,
only one degree of freedom was taken into account in the GCM calculations.
In an other work investigating the properties of $^{222}$Ra~\cite{Vauther},
we solved the CHF equations constraining both on the quadrupole and on the
octupole operators.
No GCM calculation was made, only parity projection was realized.
In a later work~\cite{Pb194}, we studied the octupole softness of the
SD state of $^{194}$Pb.
For that nucleus, we performed GCM calculations utilizing the octupole as
generating variable for different but fixed quadrupole deformation.
The octupole constraint was imposed either on $Q_{30}$, $Q_{32}$ or
a combination of these two operators.

\subsection{CHF calculations}

We are presently pursuing the latter analysis doing two-dimensional GCM
calculations for $^{194}$Pb, using the SkM$^\ast$ force.
We use the conventional $G$ pairing interaction in the BCS formalism.
To constrain on the octupole mode, we choose the operator
$q_{30}=\langle r^3Y_{30}\rangle$.
The other octupole modes which have not been included in the preesent
analysis are of two kinds.
First, the $q_{32}$ has been proven to be weakly coupled to the $q_{30}$
mode at small quadrupole deformation and negligible otherwise~\cite{Pb194}.
The odd components, $q_{31}$ and $q_{33}$, have also been recently studied
and found to generate states which lie at a higher excitation energy than
those associated to $q_{33}$~\cite{banane}.

We have considered quadrupole moments from $-40$ to $100$ barn.
This range extends well beyond the SD minimum and covers the HD region.
This is illustrated on fig.~\ref{F:echf} which shows the CHF energy as
a function of the quadrupole moment.
In addition to the usual ground state and SD minima, there exists a shallow
third minimum around $80$ b which is candidate to accomodate a HD state.
To decide whether such a state actually exists requires a dynamical
calculation.

For each value of quadrupole moment ($q_2$), we have included states with
increasing octupole deformation up to an excitation energy such that the GCM
space used in the calculation of the next section was large enough to ensure a
good accuracy on the collective spectrum and wave functions.
A maximum value of $q_{30} = 7000$ fm$^3$ was necessary to achieve this
calculation.
Fig.~\ref{F:echf} summarizes for each value of $q_{30}$ the range of $q_2$
values which has been taken into account and shows the excitation energy
domain spanned by our set of CHF basis states.

\begin{figure}[htb]
\begin{minipage}[t]{120mm}
\vskip 115mm
\caption{HFBCS deformation energy curves as a function of $q_2$ (in barn)
for values of the octupole moment ranging from 0 to 7000 fm$^3$.}
\label{F:echf}
\end{minipage}
\end{figure}

The main feature which appears on fig.~\ref{F:echf} is the softness of the SD
minimum with octupole deformation.
At $q_{30} = 0$, the SD minimum is 4.67 MeV higher than the absolute one.
At $q_{30} = 2000$ fm$^3$ it is only 1.75 MeV higher, whereas at
$q_{30} = 3000$ fm$^3$, the situation is reversed and
the second minimum is now lower than the first one by 0.9 MeV.
This feature was already present in the limited calculation of
ref~\cite{Pb194}.
A remarkable result of the present analysis is the bending of the SD valley
toward larger quadrupole deformation.
When $q_{30}$ increases from 0 to 7000 fm$^3$, the SD quadrupole moment
increases smoothly and continuously from 46 up to 58 barn.
In contrast the valley extending in the $q_{30}$ direction from the ground
state minimum is much steeper and eventually disappears around 5000 fm$^3$.
The third HD minimum located at 82 barns corresponds to an ellipsoidal
shape with a 2.1 to 1 axis ratio.
At $q_{30}=0$, the depth of this well is only 0.8 MeV.
When $q_{30}$ is increased, the quadrupole deformation of this
third minimum increases but only slightly as it smoothly disappears.
At $q_{30} = 5000$ fm$^3$, only a plateau is observed.

In the case of octupole deformation, HFBCS states with opposite octupole
moments may be used to construct eigenstates of the parity operator,
realizing thus parity projection, as was in ref.~\cite{Vauther,Pb194}.

\subsection{GCM calculations}

Once the CHF states have been constructed, we solved first the simpler
GCM equations with the quadrupole moment as a single generating variable
($q_{30}=0$).
In that case the lowest GCM state is 1.4 MeV below the absolute minimum
of the HFBCS calculation, which is a measure of the quadrupole correlations
in the ground state of $^{194}$Pb.
In all the calculations we present now, we have chosen the energy of this
state as the origin of the energy scale.
Then we solve the GCM equations with both $q_2$ and $q_{30}$ as generating
variables.
As this method has been adequately described in details~\cite{Hg92}, we will
only present here the particular aspect resulting from its generalization
to the study of the octupole mode together with the quadrupole one.

\begin{figure}[htb]
\begin{minipage}[t]{77mm}
%\framebox[74mm]{\rule[-26mm]{0mm}{80mm}}
\vskip 80mm
\caption{GCM spectrum of Pb194.
The solid line gives the HFBCS energy curve as a function of the quadrupole
moment.
Even states are marked by a vertical tickmark, odd ones by a black dot.}
\label{F:egcma}
\end{minipage}
%\end{figure}
%
\hspace{\fill}
%
%\begin{figure}[htb]
\begin{minipage}[t]{78mm}
%\framebox[74mm]{\rule[-26mm]{0mm}{80mm}}
\vskip 80mm
\caption{Enlargement of the previous figure.
States are number with increasing excitation energy.
When connected by a dotted line, they correspond to a sequence of states
generated by octupole phonon excitations.}
\label{F:egcmb}
\end{minipage}
\end{figure}

The GCM states $\mid \Psi_{k}\rangle$ result from the diagonalization
of the hamiltonian in the  nonorthogonal basis of HF+BCS states
$\mid \Phi (q^i_2,q^j_{30}) \rangle$ corresponding to different
quadrupole and octupole moments, $q^i_2$ and $q^j_{30}$,
 \begin{equation}
 \label{e102}
 \mid \Psi_{k}\rangle =
 \sum_{i,j}  f_{ij}  \mid \Phi (q^i_2,q^j_{30}) \rangle\quad   .
 \end{equation}

To solve the Hill Wheeler equations, we have to calculate the norm matrix
and the matrix elements of the hamiltonian between all the CHF states
$\mid\Phi (q^i_2,q^j_{30}) \rangle$.
As in ref~\cite{Hg92}, the mean particle number of the GCM states is
corrected by means of the Lagrange multiplier method.
Since the parity is preserved by the hamiltonian, the space of GCM solutions
is divided into positive and negative parity subspaces which do not interact.
Positive and negative parity collective wave functions and spectra were
obtained after a careful analysis of the numerical stability of the results
related to the truncation of the eigenvalues of the norm matrix on one hand,
and of the completeness of the $q_2$ -- $q_{30}$ space on the other hand.

Fig~\ref{F:egcma} shows the full GCM spectrum.
Even parity states are shown with a vertical tickmark at the mean value
of their quadrupole moment, odd states with black dots.
Due to the octupole correlations, the ground state energy gains an additional
1.4 MeV binding so that it lies 2.8 MeV below the absolute HFBCS minimum.
As compared with the GCM calculation without the octupole mode included,
all the low lying even states are shifted downward by the same 1.4 MeV energy
and have roughly the same mean quadrupole moment.
At higher  excitation energy, new even states are found.
They are be associated to excitations of octupole phonons, as we show below.

There is no state whose mean value of the quadrupole moment is close to the
third minimum although some states have components of their collective wave
function at the HD deformation.
It therefore appears that the HD well is not deep enough so that no GCM states
can qualify unambiguously as a hyperdeformed state.

\begin{figure}[htb]
\begin{minipage}[t]{160mm}
\vskip 115mm
\caption{Probability density contour plots a function the quadrupole and the
octupole moments.
First and third rows from the bottom correspond to even states,
second and fourth to odd states.}
\label{F:collwfa}
\end{minipage}
\end{figure}

Fig.~\ref{F:egcmb} shows an enlargement of fig.~\ref{F:egcma}.
States are numbered with increasing excitation energies.
They are joined in bands according to the structure of their collective wave
function as illustrated on fig.~\ref{F:collwfa} and~\ref{F:collwfb} which
show the probability density of these GCM states in the $q_2$, $q_{30}$ plane.
On fig.~\ref{F:collwfa}, one sees that the ground state collective wave
function (state 1) is localized around $q_2=q_{30}=0$, as would be an s state
in a two dimensional space.
The band (1,5,12,24) forms a sequence of states corresponding to s p d
and f waves in the $q_{30}$ direction.
The following band, next column of fig.~\ref{F:collwfa}, can be obtained from
the previous one by a quadrupole excitation.
The third and fourth bands are built on the third and fourth even GCM states
(states 3 and 4).
The fifth band, built on the eighth state, shows the limits of an
interpretation of the GCM states in the first well as being combination of
independent quadrupole and octupole elementary excitations.
The quadrupole character of this last band is more complicated than a simple
additional quadrupole excitation on top of the fourth one.
However, the sequence of octupole excitations remains visible.

\begin{figure}[htb]
\begin{minipage}[t]{130mm}
\vskip 115mm
\caption{Same as fig. 7 for GCM states at large deformations.}
\label{F:collwfb}
\end{minipage}
\end{figure}

Fig.~\ref{F:collwfb} represents three bands, the first one (left column)
is built from the SD even state (7), it also has the nice pattern of
a sequence of s, p, d and f states in the $q_{30}$ direction.
These four states are localized in the SD well.
Their mean quadrupole moment increases slightly toward the deformation of the
second well minimum.
The lowest SD even state (7) can be viewed as an s wave state located in the
SD well.
The mean quadrupole moment of state (17) corresponds to the maximum of the
barrier between the main well and the SD well.
The band which is constructed on that state, middle column of
fig.~\ref{F:collwfb}, has a much more complex structure in $q_2$.
However its structure is preserved as one adds more and more octupole phonon
excitation, although components at smaller $q_2$ are less and less important.
The last column of the figure shows a tentative assignment of states built
by octupole excitation on the most deformed state (29).
These states have significant components in the HD region, around 85 barns.
The mean value of the quadrupole moment decreases dramatically as one
adds octupole phonons.
Furthermore, these states are not at all localized in $q_2$, the dispersion
is so large that they cannot be interpreted as HD states.

We have reconstructed rotational bands on each GCM sates,
as we did in ref.~\cite{Pb194,Hgdecay}, to explore the influence of octupole
softness on the stability of SD bands.
The decay out of the band constructed on the first even SD state occurs
at the same angular momentum that was obtained already.
%The depopulation of that band happens within a couple of transitions.
Electromagnetic $E1$ and $E3$ transitions to odd bands in the first well
remain much smaller than the $E2$ transitions to the bands constructed on the
even states (1), (2), (3) and (4) so that the decay scheme of the even SD band
is not modified.

We have also constructed an odd SD rotational band on the odd state (14).
All the possible $E1$, $E2$ and $E3$ transitions to states in the first well
remain small compared to in-band $E2$ transitions.
However, $E1$ transitions to states of the even SD band are strong
so that this odd band, if it were observed, will depopulate at much higher
angular momentum than the even SD band and over a wider range of
transitions, typically between 24$\hbar$ and 36$\hbar$.
Its decay will feed the ground SD band.

\section{ROTATIONAL SUPERDEFORMED BANDS}

The Skyrme-cranking equations have been adequately described in~\cite{Mg24}.
We have chosen $^{192}$Hg as a test case to study SD rotational band
with the SkM$^\ast$ parametrization of the Skyrme force.

\subsection{Cranked Hartree Fock}

First, we have performed cranked HF calculations~\cite{Bing} without including
pairing correlations as in an earlier work for $^{24}$Mg~\cite{Mg24}.
As a function of $J_z$, the neutron et proton routhians we obtained are
quite similar to those obtained with cranked Wood-Saxon methods~\cite{RIL90},
although the assignment of Nilsson quantum numbers to our orbitals is not
always unambiguous, as discussed in~\cite{MMeyer}

Experimentally, one SD band has been observed in $^{192}$Hg~\cite{DYE90,BRH90},
identical bands have been seen in $^{194}$Pb~\cite{BRI90,THE90,HUB90} and
$^{194}$Hg~\cite{RIL90,HEN90}.
The $\gamma$ rays of the only band seen in $^{194}$Pb have energies identical
to those of $^{192}$Hg with a precision better than a couple of keV for
$12\hbar\leq J_z\leq30\hbar$.
In $^{194}$Hg three bands have been observed.
The second excited band or band 3 twins with the same $^{192}$Hg band for
15 consecutive transitions.

On the $6_{5/2}$ or the $[514]{9\over 2}$ proton orbitals,
we have built two $^{194}$Pb bands referred to below as $^{194}$Pb and
$^{194}$Pb$^\ast$.
Similarly using either the $[512]{5\over 2}$ or the $[624]{9\over 2}$
neutron orbitals we have constructed two $^{194}$Hg bands ($^{194}$Hg and
$^{194}$Hg$^\ast$ respectively).
Table~\ref{T:twin} summarizes our results at 20$\hbar$ for these five bands.

\begin{table}
\begin{minipage}[t]{158mm}
\caption{Energies, quadrupole moments, dynamical $\id$ and rigid body
${\cal J}_{\rm rig}$ (MeV$^{-1}$) moments of inertia and angular frequency
$\ho$ at 20$\hbar$ for the five SD bands studied in this work.}
\label{T:twin}
\begin{tabular}{lcccccc}
\hline
Band&E (MeV)&E$_{\rm rot}$ (MeV)&$Q_0$ (fm$^2$)&
$\id$ (MeV$^{-1}$)&${\cal J}_{\rm rig}$ (MeV$^{-1}$)&
$\ho$ (MeV) \\ \hline
$^{192}$Hg
 & -1507.734 & 1.74468 & 4446.009 & 115.407 & 119.319 & 0.17392 \\
$^{194}$Pb
 & -1513.777 & 1.81887 & 4644.611 & 110.681 & 122.365 & 0.18151 \\
$^{194}$Pb$^\ast$
 & -1511.611 & 1.73897 & 4383.145 & 116.009 & 120.133 & 0.17318 \\
$^{194}$Hg
 & -1523.683 & 1.71442 & 4393.146 & 117.925 & 120.300 & 0.17056 \\
$^{194}$Hg$^\ast$
 & -1523.350 & 1.73655 & 4446.266 & 116.279 & 120.709 & 0.17285 \\
\hline
\end{tabular}
\end{minipage}
\end{table}

The excitation energy of the $^{194}$Pb$^\ast$ band relative to $^{194}$Pb
band is 2.1MeV, while it is only 0.33MeV for the $^{194}$Hg$^\ast$.
Fig.~\ref{F:egamma} shows the differences of $\gamma$-ray energies
between the four $A=194$ bands and $^{192}$Hg.
The $\gamma$ transition energies for the $^{194}$Pb$^\ast$ and
$^{194}$Hg$^\ast$ bands differ by one to three keV from those
of $^{192}$Hg over the experimentally observed range of angular momentum.
Our calculation predicts slightly larger
differences at higher angular momentum ($J_z\geq32\hbar$).
In contrast, the $^{194}$Pb and $^{194}$Hg bands do not resemble each
other nor the $^{192}$Hg band.
By discrete differentiation of $J_z$ with respect to $\ho$,
we have calculated the dynamical moment of inertia $\id$ of these bands
as a function of $J_z$.
For the five bands the HF value of $\id$ grows steadily with $J_z$.
This result differs from those of other studies based on either
Woods-Saxon~\cite{CHA89} or rotating oscillator~\cite{RAG90} potentials
in which pairing effects have also been neglected.
The values of the HF moments of inertia disagree with the observed ones;
they are too large for small $J_z$'s and too small at the
largest angular momentum $J_z=40\hbar$.
This indicates that data can only be fully understood
when the variation of pairing correlations with rotation
is correctly taken into account.

\begin{figure}[htb]
\begin{minipage}[t]{130mm}
%\framebox[159mm]{\rule[-26mm]{0mm}{140mm}}
\vskip 95mm
\caption{Differences of $\gamma$-ray energies between the lowest SD band of
$^{192}$Hg and the four $A=194$ SD bands: $^{194}$Pb ($+$), $^{194}$Pb$^\ast$
($\times$), $^{194}$Hg ($\triangle$) and $^{194}$Pb$^\ast$ ($\Diamond$).
The horizontal dashed lines indicate the $\pm2$keV accuracy, as a measure of
band identity.}
\label{F:egamma}
\end{minipage}
\end{figure}

Table~\ref{T:twin} shows that the variations of the quadrupole moments
from one band to the other are not related to the changes in dynamical
moments of inertia.
The $^{194}$Pb band has a larger quadrupole moment and a smaller
moment of inertia while the situation is reversed for the $^{194}$Hg band.
In addition, the HF predicted behavior of $\id$ versus $J_z$
is contrary to that of the
mass quadrupole moment $Q_0$ which decreases continuously
(although by less than 5\%) when the angular momentum grows.
Neither the dynamical moment of inertia nor
the quadrupole moments obey a simple $A^{5/3}$ rule.
As expected the rigid moments of inertia follow
the same trend as the quadrupole moments. They are always significantly
different from the $\id$ values.

As was shown in  other works ~\cite{RAG90,DSW92}, our calculations indicate
that twinning is related to the filling of specific single-particle orbitals.
However our calculations prove that the occupation of these ``twinning''
orbitals does not generate a modification of those bulk dynamical properties
related to rotation.
A full understanding of the microscopic self-consistent origin of this
rather delicate balance mechanism will require a deeper analysis of the
structure of these orbitals and of their interaction with the mean-field.

\subsection{Cranked HFB equations -- preliminary results}

Pairing correlations are becoming weaker as the angular momentum increases.
They may be expected to be small in the rare earth region where SD
bands are not observed at low spins, although this conjecture remains
to be verified quantitatively.
In the Hg region, even though these bands are not observed at zero
angular momentum, they extend to much lower spin,
presumably below 10$\hbar$.
For this reason, it is necessary to extend the cranked HF mean field
so as to include pairing.
Because rotation breaks time-reversal symmetry, the BCS approximation
is not valid: it is not possible to assume that pairs are built
on only two orbitals of the HF mean-field;
one cannot escape the full complexity of the HFB equations.

The implementation of the Bogoliubov transformation in the cranked CHF
scheme can be summarized as follows (see ref.~\cite{tokyo}):
\begin{itemize}
\item Given the self-consistent HF mean-field in coordinate space $\hf$,
find the eigenvalues and eigenfunctions of $\hf-\oj$
in both positive and negative signature spaces.
\item Calculate the relevant matrix elements of the two-body pairing
interaction in the basis which diagonalizes $\hf-\oj$.
Construct the gap matrix and solve the HFB equations within
the space of positive signature.
 From the corresponding eigenvalues and the eigenvectors, deduce the HFB
matrices $U$ and $V$ and the quasiparticle energies $E$ for both signatures.
\item Using the matrices $U$ and $V$, construct the density matrix
and the pairing tensor.
\item A diagonalization of the density matrix provides occupation numbers
and the eigenfunctions (distinct from the eigenvalues of $\hf-\oj$, as
$\rho$ and $\hf$ do not commute).
 From them, one constructs the density matrix in coordinate space which is
then used to build the HF hamiltonian $\hf$ associated with the Skyrme force.
\end{itemize}

This set of equations is solved self-consistently with a separable pairing
force which reduces to the seniority interaction when the angular velocity
$\omega$ vanishes.
Its neutron and proton strengthes have been adjusted to reproduce the
static pairing properties of nuclei in the neutron deficient Hg-Pb area
of the isotope map.

We have calculated the SD lowest band of $^{192}$Hg from $J_z=0$ to 44$\hbar$.
Several features appear in the calculation:
i) the neutron and proton pairing correlations which are introduced in this HFB
calculation disappear at 24$\hbar$ and 40$\hbar$ respectively.
ii) the difference between the total HF and HFB energies is smaller than the
sum of neutron and proton pairing energies because there is a loss of binding
energy in the mean-field contribution.
iii) the ratio of the proton to neutron contributions to angular momentum is
always less than $Z/N$; as $\jz$ grows, it decreases since the superfluidity
of neutrons is the first to disappear then it increases smoothly to reflect
the fast decrease of proton pairing correlations to finally reach the HF value.

\begin{figure}[htb]
\begin{minipage}[t]{140mm}
\vskip 140mm
\caption{Dynamical moment of inertia in MeV$^{-1}$ as a function of $\ho$.}
\label{F:inertia}
\end{minipage}
\end{figure}

Figure~\ref{F:inertia} compares the experimental dynamical moment of inertia
to the calculated one within HFB as a function of $\ho$.
At zero frequency, the HFB moment of inertia, $\id = 80$ MeV$^{-1}$, is
slightly smaller than the experimental extrapolation, in contrast with the HF
calculation without pairing of ref.~\cite{Bing}, $\id = 110$ MeV$^{-1}$.
Then $\id$ increases with $\ho$ much faster than experiment.
The disappearance of the neutron pairing correlations at 24 $\hbar$
is reflected on the HFB $\id$ by an abrupt reduction.
Similarly, when the proton pairing vanishes at 40$\hbar$, the HFB moment of
inertia reduces to the HF one.
The fast increase of the HFB $\id$ between 0 and 24 $\ho$ indicates that
pairing correlations do not decreases fast enough with increasing frequency.
The sharp reduction at 24 and 40 $\hbar$ is due to their too early and too
sudden disappearance.

Rather than discussing the pairing correlations themselves, it seems natural
to look for deficiencies in the mean-field (HFB) treatment of these
correlations as the nucleus is a finite system.
A well established way to improve the HFB description is provided by the
variation after projection (VAP) method on the correct nucleon number.
This method leads to heavier but still tractable calculations.
 From calculations~\cite{RS80} performed for other nuclei and at smaller
deformations, VAP is known to suppress the unphysical sharp HFB phase
transition and to maintain pairing correlations to higher angular momentum
while softening their decrease.

In view of the present results, we have included in the HFB equations the
approximate VAP realized by the LN prescription~\cite{Lipnog}.
The resulting dynamical moment of inertia is shown on fig.~\ref{F:inertia},
it rises continuously from 0 to 30 $\hbar$, angular momentum at which we
ended this calculation.
As expected, neutron and proton pairing decreases smoothly, but do not
vanishes.
At 30$\hbar$, the proton pairing energy is 50\% what it is at 0$\hbar$,
whereas for neutron it is 40\%.
As the rise in $\id$ is too high as compared with experiment, we have
arbitrarily decrease strength of the pairing interaction by 10\% and repeated
our HFB+LN calculation up to higher $\ho$, reaching an angular momentum of
46$\hbar$.
As can be seen from fig.~\ref{F:inertia}, the overall behavior of $\id$ agrees
rather well with experiment from $\ho=0$ to 300 keV, however, it fails to
reproduce experiment at higher frequency.

Since the dynamical moment of inertia is defined as $(d\ho/d\jz)^{-1}$ one
expects $\id$ to grow until it reaches a maximum and then to decrease to the
HF value.
This behavior is very much alike that found in other
analyses~\cite{CHA90,MOO90}
both qualitatively and quantitatively.
However it is markedly different from the data~\cite{POR92} which shows a
steadily increasing moment of inertia up to the largest observed value of
$\ho$ which is above 400 keV.

\section{CONCLUSIONS}

In this review, we have focused our attention on our work and showed the
the predictive power of mean field calculations complemented by the GCM
and projection, starting from an effective force of the Skyrme type.
For that purpose, we have selected as an illustration some topics, namely
the isotope shifts in $^{194}$Pb and the effective force, quadrupole and
octupole coupling in $^{194}$Pb SD collective states and finally, rotation
without and with pairing correlations included.
Many more physical situations have been or can be also studied.
Let us mention only few of them such as a successful analysis of the odd-even
staggering in the Pb isotope shift~\cite{Taj93}, a study of ``banana''
modes~\cite{banane} or a detailed calculation of the electric
decay rate of some fission isomers~\cite{Kri.ThU}.

Along the same path, there are other works which in different ways achieve
the same goal.
For example, instead of Skyrme forces, the finite range D1S
interaction~\cite{D1S} has been extensively used.
However the GCM equations are not solved exactly, rather the different groups
utilizing D1S make use of the Gaussian Overlap Approximation which yields
a collective Bohr Hamiltonian.
For instance, SD bands have been analyzed in the Hg region~\cite{BIII}.
Still with D1S, the Madrid group has calculated spectroscopic properties of
nuclei ranging from the rare earth region~\cite{Erbium} to Xenon
nuclei~\cite{Xenon}.
Let us also mention a recent work where they have projected their collective
wave function onto good angular momentum to calculate transition
probabilities~\cite{Madrid}.
Finally relativistic mean field approach~\cite{Serot,Ring90} has been
applied to calculate nuclear properties.
Results obtained so far are quite promising~\cite{Ring92}.

%\newpage


\vskip 4em

\begin{thebibliography}{9}
\bibitem{Zrpaper} P. Bonche, H. Flocard, P.H. Heenen, S.J. Krieger
                  and M.S. Weiss, Nucl. Phys. {\bf A443} (1985) 39
\bibitem{PB89}    P. Bonche, S.J. Krieger, Ph. Quentin, M.S. Weiss, J. Meyer,
                  M. Meyer, N. Redon, H. Flocard and P.H. Heenen,
                  Nucl. Phys. {\bf A500} (1989) 308
\bibitem{Isthmus} S.J. Krieger, P. Bonche, M.S. Weiss, J. Meyer, H. Flocard
                  and P.H. Heenen, Nucl. Phys. {\bf A542} (1992) 43
\bibitem{Mey93}   J. Meyer et al. in preparation
\bibitem{Mg24}    P. Bonche, H. Flocard and P.H. Heenen,
                  Nucl. Phys. {\bf A467} (1987) 115
\bibitem{Hg92}    P. Bonche, J. Dobaczewski, H. Flocard, P.H. Heenen
                  and J. Meyer, Nucl. Phys. {\bf A510} (1990) 466
\bibitem{Proj}    P.H. Heenen, P. Bonche, J. Dobaczewski and H. Flocard,
                  Nucl. Phys. in press
\bibitem{Srpaper} P. Bonche, J. Dobaczewski, H. Flocard and P.H. Heenen,
                  Nucl. Phys. {\bf A530} (1991) 149
\bibitem{Lipnog}  H.J. Lipkin, Ann. Phys. (New-York) {\bf 9} (1960) 272\\
                  Y. Nogami, Phys. Rev. {\bf 134} (1964) B313
\bibitem{Pb1s}    L. Bennour, P.H. Heenen, P. Bonche, J. Dobaczewski and
                  H. Flocard, Phys. Rev. {\bf C40} (1989) 2834
\bibitem{Taj93}   N. Tajima, P. Bonche, H. Flocard, P.H. Heenen and
                  M.S. Weiss, Nucl. Phys. {\bf A551} (1993) 434
\bibitem{Bing}    B.Q. Chen, P.H. Heenen, P. Bonche, M.S. Weiss and
                  H. Flocard, Phys. Rev. {\bf C46} (1992) 1582
\bibitem{Isoexp}  E.W. Otten, Nuclear radii and moments of unstable nuclei,
                  Treatrise in heavy ion science, vol. 8, ed D.A. Bromley
                  (Plenum, New York, 1988) p. 515
\bibitem{SIII}    M. Beiner, H. Flocard, N. Van Giai and Ph. Quentin,
                  Nucl. Phys. {\bf A238} (1975) 29
\bibitem{Skmast}  J. Bartel, Ph. Quentin, M. Brack, C. Guet
                  and H.-B. H\aa kansson,
                  Nucl. Phys. {\bf A386} (1982) 79
\bibitem{Sg2}     N. Van Giai and H. Sagawa, Phys. Lett. {\bf B106} (1981) 379
\bibitem{SkM}     H. Krivine, J. Treiner and O. Bohigas,
                  Nucl. Phys. {\bf S336} (1980) 155
\bibitem{Jfb}     J.F. Berger, private communication.
\bibitem{D1}      D. Gogny, in Proc Int. Conf. on Nuclear Physics,
                  eds J. De Boer and H. Mang, (North Holland, Amsterdam, 1973)
\bibitem{D1S}     J. Decharg\'e, M. Girod and D. Gogny,
                  Phys. Lett. {\bf 55B} (1975) 361\\
                  J. Decharg\'e and D. Gogny, Phys. Rev. {\bf C21} (1980) 1568
\bibitem{Ma80}    J. Martorell and D.W.L. Sprung,
                  Z. Phys. {\bf A298} (1980) 153
\bibitem{Tho83}   R.C. Thompson, M. Anselment, K. Bekk, S. G\"{o}ring,
                  A. Hanser, G. Meisel, H. Rebel, G. Schatz, and B.A. Brown,
                  J. of Phys. {\bf G9} (1983) 443
\bibitem{Din87}   U. Dinger, J. Eberz, G. Huber, R. Menges, S. Schr\"{o}der
                  R. Kirchner, O. Keppler, T. K\"{u}hl, D. Marx and
                  G.D. Sprouse, Z. Phys. {\bf A328} (1987) 253
\bibitem{Tondeur} F. Tondeur, Nucl. Phys. {\bf A315} (1979) 353
\bibitem{Kri90}   S.J. Krieger, P. Bonche, H. Flocard. Ph. Quentin and
                  M.S. Weiss, Nucl. Phys. {\bf A517} (1990) 275
\bibitem{Landau}  J. Speth, E. Werner and W. Wild,
                  Phys. Rep. {\bf C33} (1977) 127
\bibitem{Cha76}   R.R. Chasman, Phys. Rev. {\bf C14} (1976) 1935
\bibitem{BE91}    G.F. Bertsch and H. Esbensen,
                  Ann. Phys. {\bf 209} (1991) 327
\bibitem{Mosko}   I.M. Green and S.A. Moszkowski,
                  Phys. Rev. {\bf139} (1965) B790
\bibitem{Wiringa} R.B. Wiringa, V. Fiks and A. Fabrocini,
                  Phys. Rev. {\bf C38} (1998) 1010
\bibitem{FP}      B. Friedman and V.R. Pandharipande,
                  Nucl. Phys. {\bf A361} (1981) 502
\bibitem{Chabana} E. Chabanat et al. in preparation
\bibitem{Vauther} P. Bonche, P.H. Heenen, H. Flocard and D. Vautherin,
                  Phys. Lett. {\bf 175B} (1986) 387
\bibitem{Pb194}   P. Bonche, S.J. Krieger, M.S. Weiss, J. Dobaczewski,
                  H. Flocard and P.H. Heenen,
                  Phys. Rev. Lett. {\bf 66} (1991) 876
\bibitem{banane}  J. Skalski, P.H. Heenen, P. Bonche, H. Flocard and J. Meyer,
                  Nucl. Phys. {\bf A551} (1993) 109
\bibitem{Hgdecay} P. Bonche, J. Dobaczewski, H. Flocard, P.H. Heenen,
                  S.J. Krieger, J. Meyer and M.S. Weiss,
                  Nucl. Phys. {\bf A519} (1990) 509
\bibitem{RIL90}   M.A. Riley et al. Nucl. Phys. {\bf A512} (1992) 178
\bibitem{MMeyer}  M. Meyer et al. Phys. Rev. {\bf C45} (1992) 233
\bibitem{DYE90}   D. Ye et al. Phys. Rev. {\bf C41} (1990) R13
\bibitem{BRH90}   J.A. Becker et al. Phys. Rev. {\bf C41} (1990) R9
\bibitem{BRI90}   M.J. Brinkman et al. Z. Phys. {\bf A336} (1990) 115
\bibitem{THE90}   K. Theine et al. Z. Phys. {\bf A336} (1990) 113
\bibitem{HUB90}   H. H\"ubel et al. Nucl. Phys. {\bf A520} (1990) 125c
\bibitem{HEN90}   E.A. Henry et al. Z. Phys. {\bf A335} (1990) 361
\bibitem{CHA89}   R.R. Chasman, Phys. Lett. {\bf B219} (1989) 227
\bibitem{RAG90}   I. Ragnarsson, Nucl. Phys. {\bf A520} (1990) 67c
\bibitem{DSW92}   J. Dudek, Z. Szimanski and T. Werner, to be published
\bibitem{tokyo}   H. Flocard, B.Q. Chen, B. Gall, P. Bonche, J. Dobaczewski,
                  P.H. Heenen and M.S. Weiss,
                  Nucl. Phys. {\bf A557} (1993) 559c
\bibitem{RS80}    P.~Ring and P.~Schuck, The Nuclear Many-Body Problem,
                  Springer Verlag, (1980)
\bibitem{CHA90}   R.R. Chasman, Phys. Lett. {\bf B242} (1990) 317
\bibitem{MOO90}   E.F. Moore et al., Phys. Rev. Lett. {\bf 64} (1990) 3127
\bibitem{POR92}   M.G. Porquet, private communication
\bibitem{Kri.ThU} S.J. Krieger et al., unpublished
\bibitem{BIII}    M. Girod, J.P. Delaroche, J. Libert and I. Deloncle,
                  Phys. Rev. {\bf C45} (1992) R1420\\
                  J. Libert, J.F. Berger, J.P. Delaroche and M. Girod,
                  Nucl. Phys. {\bf A553} (1993) 523c
\bibitem{Erbium}  J.L Egido and L.M. Robledo,
                  Phys. Rev. Lett. {\bf 70} (1993) 2876
\bibitem{Xenon}   V. Martin and L.M. Robledo, Phys. Rev. C, in press
\bibitem{Madrid}  J.L Egido, L.M. Robledo and Y. sun,
                  Nucl. Phys. {\bf A560} (1993) 253
\bibitem{Serot}   B.D. Serot and J.D. Walecka,
                  Adv. Nucl. Phys. {\bf 16} (1986) 1
\bibitem{Ring90}  Y.K. Gambhir, P. Ring and A. Thimet
                  Ann. Phys. (N.Y.) {\bf 198} (1990) 132
\bibitem{Ring92}  J.P. Maharana, Y.K. Gambhir, J.A. Sheikh and P. Ring,
                  Phys. Rev. {\bf C46} (1992) R1163
%\bibitem{LAU92}   T. Lauritsen et al. Phys. Lett. B 279 (1992) 239




\end{thebibliography}
\end{document}